\documentclass[12pt]{article}

\usepackage{amssymb}
\usepackage{amsmath}
\usepackage{graphbox}
\usepackage{graphicx}

\begin{document}

\begin{titlepage}

\begin{flushright}
arXiv:1903.10100
\end{flushright}
\vskip 2.5cm

\begin{center}
{\Large \bf There is No Ambiguity in the Radiatively Induced
Gravitational Chern-Simons Term}
\end{center}

\vspace{1ex}

\begin{center}
{\large Brett Altschul\footnote{{\tt baltschu@physics.sc.edu}}}

\vspace{5mm}
{\sl Department of Physics and Astronomy} \\
{\sl University of South Carolina} \\
{\sl Columbia, SC 29208} \\
\end{center}

\vspace{2.5ex}

\medskip

\centerline {\bf Abstract}

\bigskip

Quantum corrections to Lorentz- and CPT-violating QED in flat spacetime produce
unusual radiative corrections, which can be finite but of undetermined magnitude.
The corresponding radiative corrections in a gravitational theory are
even stranger, since the term in the fermion action involving a preferred
axial vector $b^{\mu}$ would give
rise to a gravitational Chern-Simons term that is proportional $b^{\mu}$, yet
which actually does not break Lorentz invariance. Initially, the coefficient of this
gravitational Chern-Simons term appears to have the same ambiguity as the
coefficient for the analogous term in QED. However, this puzzle is resolved
by the fact that the gravitational theory has more stringent gauge invariance
requirements. Lorentz symmetry in a metric theory of gravity can only be broken
spontaneously, and when the vector $b^{\mu}$ arises from spontaneous symmetry
breaking, these specific radiative corrections are no longer ambiguous but instead must
vanish identically.

\bigskip

\end{titlepage}

\newpage

\section{Introduction}

Since the 1990s, there had been a significant renewal of interest in the possibility
that the seemingly absolute Lorentz and CPT symmetries of the standard model and
gravity might actually be very weakly violated. At this time, there is no
no compelling evidence for such symmetry breaking. However, if violations of
isotropy, Lorentz boost invariance, or CPT were ever observed experimentally, the
discovery would obviously be of fundamental importance. It would change our
understanding of how physics works at the very deepest levels. Even if there is
no current evidence for Lorentz or CPT violation, these symmetries are so basic
(as fundamental building blocks of both quantum field theories and the general
theory of relativity), that they are worthy of careful study.

There are also other reasons to be interested in Lorentz and CPT tests. Attempts
to develop a quantum theory of gravity have shown that many of the 
speculative frameworks that have been suggested to describe quantum gravity
seem to allow for the existence of Lorentz violation, at least within certain regimes.
Moreover, with the explication of a comprehensive effective field theory (EFT)
capable of describing Lorentz-violating phenomena, it came to be realized that
the symmetry violations could come in a much wider variety of forms than previous
unsystematic analyses had considered. Large regions of the EFT parameter space
were scarcely constrained by earlier generations of experiments. CPT symmetry is
also closely tied to Lorentz symmetry, so that even with non-local interactions,
CPT violation in a quantum field theory (QFT) automatically entails Lorentz
violation~\cite{ref-greenberg}, as long as the theory has a well-defined
$S$-matrix.

The general EFT describing Lorentz violation in particle physics therefore includes
the most general CPT violation as well. This EFT is known as the standard model
extension (SME)~\cite{ref-kost1,ref-kost2}. The action for the SME contains
operators that can be constructed out of the standard model fields.
The usual standard model action is formed by writing down all the local, renormalizable,
$SU(3)_{c}\times SU(2)_{L}\times U(1)_{Y}$
gauge-invariant, Lorentz-invariant operators that can be constructed from those fields.
The SME is constructed in much the same way, but with the Lorentz invariance requirement
dropped; specifically, the minimal SME (mSME) keeps all of the other requirements---locality,
renormalizability, and gauge invariance. The mSME is now the usual framework
used for parametrizing the results of experimental Lorentz and CPT
tests. However, since the mSME action contains quite a large number of parameters,
many different types of experiments have turned out to be useful for establishing
bounds on the mSME parameters. An up-to-date summary of the results of these
experiments is given in~\cite{ref-tables}.

Studies of possible Lorentz and CPT violation have also been fruitful theoretically,
providing new insights, especially into the structure of QFTs.
The radiative corrections to the Chern-Simons term in Lorentz-violating
quantum electrodynamics (QED) have been one of the most studied topics
related to the SME---and almost certainly the most controversial.

There is a similarly-structured gravitational Chern-Simons term in the gravitational version
of the SME. Although the radiative corrections in the gravitational sector have been
examined to a limited extent, a profound and significant puzzle exists in that
sector---which has previously been overlooked. This paper will both introduce
this puzzle and proceed to solve it.

\section{The Puzzle of Gravitational Radiative Corrections}

The basic outline of the puzzle is the following. We must begin with a discussion of the
simpler electromagnetic Lorentz violation term. The Chern-Simons term in the (3+1)-dimensional
Abelian gauge sector takes the form ${\cal L}_{AF}=\frac{1}{2}k_{AF}^{\mu}\epsilon_{\mu\alpha\beta\gamma}
F^{\alpha\beta}A^{\gamma}$~\cite{ref-carroll1}. (When $k_{AF}^{\mu}$ is purely timelike, this term
in the Lagrange density is proportional to $\vec{A}\cdot\vec{B}$, which breaks P and CPT symmetries.)
Coming from the charged fermion sector,
there is a radiative correction to ${\cal L}_{AF}$ that is necessarily finite, but whose coefficient
is undetermined. At the quantum level, there is an infinite family of
different theories that correspond to the same classical Lagrangian. The
differences between these theories are in how they are regulated, but there is
nothing that singles out one regulator as being unambiguously correct. Different high-momentum regulators
lead to  radiatively-generated terms with different finite
coefficients~\cite{ref-coleman,ref-jackiw1,ref-victoria1,ref-chung1,ref-chung2,ref-chen,ref-chung3,
ref-volovik,ref-chan,ref-bonneau1,ref-chaichian,
ref-victoria2,ref-battistel,ref-andrianov,ref-altschul1,ref-altschul2}.
Various schemes have been suggested for identifying a single
correct result; some authors have argued
that only their certain specific regulators were appropriate for the calculation---and thus
that there was one true correct answer.  
However, all such unambiguous answers appear to suffer from one
of two deficiencies. The most naive schemes fix the term by demanding that
the Lagrange density be gauge invariant; since the Chern-Simons term is not
gauge invariant (it changes by a total derivative under a gauge transformation),
the term is automatically ruled out. However, this is not a legitimate
result, because it excludes the term of interest {\em a priori}. Gauge
invariance of the Lagrange density is an unnecessarily strong condition; if
we instead only demand that the integrated action be invariant, the
Chern-Simons term is fully allowed. Alternatively, nonperturbative schemes for
fixing the radiative correction have also been suggested. However, for a
nonperturbative framework to make sense, it must provide a way of
determining higher-order radiative corrections as well as first-order ones; and unfortunately,
all the proposed nonperturbative methodologies that lead to particular nonzero values of
the induced Chern-Simons coefficient appear to fail at higher order.

Calculations appear to show that the gravitational sector has the same
kind of ambiguity~\cite{ref-mariz2,ref-mariz1,ref-gomes1,ref-filipe,ref-assuncao}.
The Lorentz-violating Chern-Simons term for (3+1)-dimension gravity is
\begin{equation}
{\cal L}_{\Gamma}=-\frac{1}{4}v_{\mu}\epsilon^{\mu\alpha\beta\gamma}\left(
\Gamma_{\alpha\tau}^{\sigma}\partial_{\beta}\Gamma_{\gamma\sigma}^{\tau}
+\frac{2}{3}\Gamma_{\alpha\tau}^{\sigma}\Gamma_{\beta\eta}^{\tau}\Gamma_{\gamma\sigma}^{\eta}
\right),
\end{equation}
in terms of the Christoffel symbols $\Gamma_{\alpha\beta}^{\gamma}$
[and with $\kappa=(8\pi G)^{-1}=1]$. In the linearized
gravity limit, this may be expressed more conveniently directly in terms of the metric fluctuations.
${\cal L}_{\Gamma}$ is proportional to $v^{\mu}\epsilon_{\mu\alpha\beta\gamma}h^{\beta\nu}
\partial^{\gamma}(\partial_{\sigma}\partial^{\sigma}h^{\alpha}_{\nu}-\partial_{\nu}
\partial_{\sigma}h^{\alpha\sigma})$, where $g^{\mu\nu}=\eta^{\mu\nu}+h^{\mu\nu}$.
Not surprisingly, the gravitational Chern-Simons term contains two more derivatives than
the electromagnetic term (to match the number
of free $h^{\mu\nu}$ indices); but otherwise the two types of terms appear (in the weak field limit) to be
quite similar in structure. However, there is actually a fundamental difference between the
electromagnetic Chern-Simons term that may be radiatively generated in
Lorentz-violating QED and the corresponding gravitational Chern-Simons term.
The difference is that the gravitational Chern-Simons term is, in spite of appearances,
actually Lorentz invariant. The profound puzzle that faces us is that it appears to be
possible for a Lorentz-violating $b^{\mu}$ term in the fermion sector to generate
a radiative correction that is proportional to $b^{\mu}$ and fully
P violating~\cite{ref-alexander1,ref-smith,ref-alexander2}, yet which is invariant
under all rotations and Lorentz boosts.

The Lorentz invariance of a pure gravity theory that includes a Chern-Simons
term is rather subtle. In fact, this was itself a bit of a puzzle when the term was
first introduced~\cite{ref-jackiw5}; it appeared that there were no physical
distinctions between versions of the theory with explicit (externally imposed)
symmetry breaking and certain types of dynamical symmetry breaking. However, this was ultimately explained, and
the Lorentz symmetry of the gravitational Chern-Simons theory was demonstrated
by constructing the conserved gravitational energy-momentum (pseudo-)tensor
$\Theta^{\mu\nu}$~\cite{ref-guarrera}. This $\Theta^{\mu\nu}=\Theta^{\nu\mu}$
has a symmetric form, and symmetry of the energy-momentum tensor is equivalent to Lorentz
invariance of the $S$-matrix (because the rotation and boost generators can be expressed as
integrals of moments of $\Theta^{\mu\nu}$).
Evidently, the dependence of the theory on the preferred vector
$v^{\mu}$ is illusory. It is not possible to write down such a Chern-Simons
term without introducing such a vector, but the particular spacetime direction of $v^{\mu}$
turns out to have no bearing on the physics. This is directly related to
the gauge invariance of the theory; the semblance of Lorentz violation is essentially a
gauge artifact.

Nonetheless, the gravitational Chern-Simons term really does break the
discrete symmetries of general relativity. For a timelike $v^{\mu}$, the boost
violation that is seemingly apparent in the form of the term is unphysical, but
the parity violation is quite real. Boost invariance manifests itself in the fact
that all gravitational waves in the theory propagate at the speed of light. However,
the P breaking means that right- and left-polarized waves are coupled
to their sources with different strengths. Note that the lack of Lorentz
violation in the CPT-violating gravitational Chern-Simons theory demonstrates
that CPT violation in the gravitational sector does not automatically need to be accompanied by
Lorentz violation. This is a somewhat surprising result, although it is clear upon
careful reinspection
that the formal derivation~\cite{ref-greenberg} of the result that CPT violation
requires Lorentz violation does not technically apply in the context of a metric
theory of gravity.
Purely gravitational theories are not formulated using QFT to begin with, and
stability of the quantum vacuum (required for the definition of the $S$-matrix)
is thus not a condition that can formally be applied.
Physically, this corresponds to the fact that there is generally
no reason to expect that matter in a universe cannot progressively coalesce into heavier and
heavier black holes, there being no ultimately stable lowest-energy configuration.

So while $b^{\mu}$ and $v^{\mu}$ terms have the same discrete symmetries, $b^{\mu}$ breaks
Lorentz invariance, while $v^{\mu}$ does not. Returning to the main problematical
observation, it appears that if a fermion species with a Lorentz-violating $b^{\mu}$
term is coupled to gravity, radiative corrections may produce a ${\cal L}_{\Gamma}$ term with
coefficient $v^{\mu}$ proportional to $b^{\mu}$. The radiative correction would thus
possess a much greater degree of symmetry than the novel term that generated
it. It is not immediately clear whether this is possible or whether it should
be ruled out by some general principles of field theory. Whichever option is
correct, there is evidently quite a bit more to be understood about how 
radiative corrections work in these kinds of theories.

Having established the existence of this open question,
we shall show that the resolution of this enigma is rather subtle. The
key pieces of information necessary to construct the solution are embedded in the
theory, but they need to be pieced together, in conjunction with what is
already known about the general structure of Lorentz-violating field theories.
The ultimate answer will be tied to the fact that gravitational theories
are fundamentally different from other field theories when it comes to
Lorentz violation. In particular, Lorentz violation in a metric theory must arise
spontaneously.

\section{Structure of Ambiguous Corrections}

The fact that Lorentz symmetry in a metric theory of gravity can only be broken
spontaneously will have profound consequences for the radiative corrections to the
gravitational Chern-Simons term. To understand these consequences, we must look very
carefully at the structures of both the electromagnetic and gravitational Chern-Simons
terms. This will reveal a close connection to chiral anomalies.

\subsection{Abelian Theory}

\begin{figure}
\centering
\includegraphics{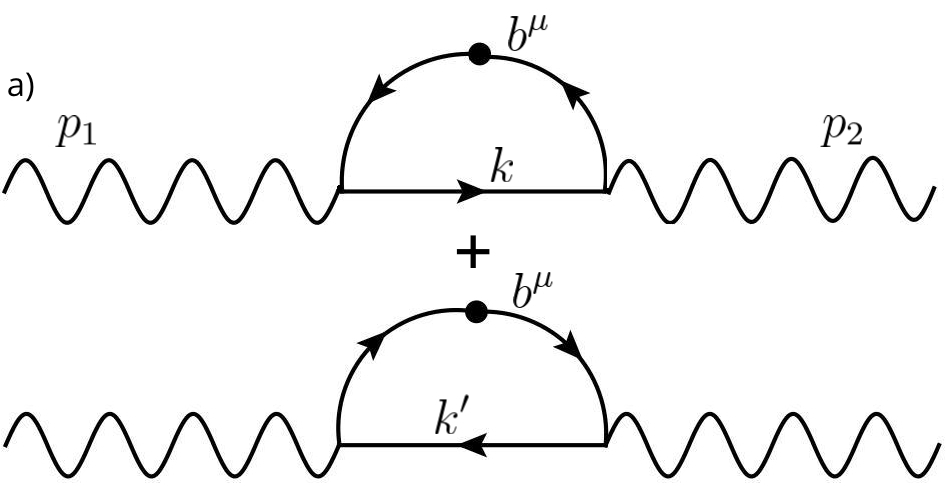}
\includegraphics{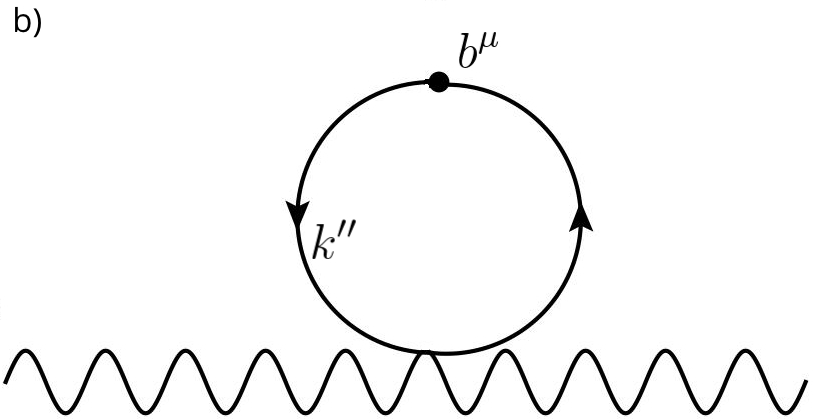}
\caption{One-loop diagrams that can contribute to the radiatively-generated
Lorentz-violating Chern-Simons terms.
The dots represent the $b^{\mu}$ insertions appearing in the fermion propagator
$S_{b}$. (a) The two triangle
diagrams that exist in the radiative calculation of the Abelian Chern-Simons term.
(b) The additional contributing diagram that appears in the gravitational theory.
\label{fig-contribute}}
\end{figure}

The QED Lagrange density, including the only mSME term that has the
right structure to make a radiative contribution to ${\cal L}_{AF}$, is
\begin{equation}
{\cal L}_{{\rm QED}}=-\frac{1}{4}F^{\mu\nu}F_{\mu\nu}+\bar{\psi}(i\!\!\not\!\partial-m-
q\!\!\not\!\!A\,+\!\not\!b\gamma_{5})\psi.
\end{equation}
The pair of diagrams that contribute to the undetermined radiative correction in an
Abelian gauge theory is shown in Figure~\ref{fig-contribute}a. These are
essentially just the same dia\-grams---with fermion triangles and two external photons---that
are responsible for the chiral anomaly.

The two diagrams in Figure~\ref{fig-contribute}a differ in the direction of the fermion number flow around the
triangular loop. Alternatively (taking the viewpoint suggested by the
nonperturbative treatment in~\cite{ref-jackiw1}), there is just a single
one-loop diagram---the usual vacuum polarization diagram, but with the modified
fermion propagator
\begin{equation}
S_{b}(l)=\frac{i}{\!\not l-m\,+\!\not\!b\gamma_{5}}
\approx\frac{i}{\!\not l-m}+\frac{i}{\!\not l-m}\left(i
\!\not\!b\gamma_{5}\right)\frac{i}{\!\not l-m}.
\end{equation}
(Since the Lagrange density involves no nonstandard time derivatives, the fermion sector
may be quantized without any changes to the spinor representation~\cite{ref-kost3,ref-kost5},
and the $b^{\mu}$-exact propagator may simply be read off from ${\cal L}_{{\rm QED}}$.)
The two triangles then arise from the fact that, at first order in the Lorentz violation, there
may be a $b^{\mu}$ insertion on exactly one of the two internal fermion lines.

The two triangle diagrams are very similar to those that arise in the
calculation of the QED chiral anomaly---for example, in its original context of
$\pi^{0}$ decay~\cite{ref-bell1}. Each triangle has two vertices attached to outgoing gauge
boson propagators, and a third axial vector vertex with a $\gamma_{5}$. The presence of the
$\gamma_{5}$ is what ensures that, when the loop momentum is very large, the
contributions from the two different diagrams cancel out, since the fields
passing through the $\gamma_{5}$ vertex have opposite chiralities. A great deal
is known about the structure of these kinds of diagrams, from analyses of the
chiral anomaly.

However, there is a subtle but fundamental difference between
how the sum of the two triangle diagrams should be evaluated, in the
contexts of Lorentz violation versus $\pi^{0}$ decay. The issue is that, when
a meson vertex is involved, there is an additional leg attached there, which
represents the incoming decaying particle. That particle carries momentum,
so the two fermion propagators attached to that vertex will have different momenta.
In contrast, in a theory with explicit Lorentz violation, the $b^{\mu}$ vertex cannot
carry any momentum whatsoever, because $b^{\mu}$ is constant across all spacetime.
Surprisingly, this changes the way the calculations can proceed in a significant way.

It was initially argued~\cite{ref-coleman}---incorrectly---that
the triangle diagrams could not generate
a Chern-Simons term, because of the theory's gauge invariance properties.
In each diagram, there are external photons attached to two of the triangle's corners, carrying
momenta $p_{1}$ and $p_{2}$. Ward identities then imply that the amplitude ${\cal M}_{\mu\nu}$
corresponding to the sum of the two fermion loops must be transverse to both $p_{1}$ and
to $p_{2}$. That is,
\begin{equation}
\label{eq-MWard}
{\mathcal M}_{\mu\nu}p_{1}^{\mu}={\mathcal M}_{\mu\nu}p_{2}^{\nu}=0.
\end{equation}
The implies that the amplitude must be ${\cal O}(p_{1})$ and separately ${\cal O}(p_{2})$.
If $p_{1}$ and $p_{2}$ are allowed to be different (that is, if the axial vector vertex can
carry a nonzero momentum), this implies that ${\mathcal M}_{\mu\nu}$ is ${\mathcal O}(p_{1}p_{2})$.
When we set $p_{1}=-p_{2}$, corresponding to the physical situation, the amplitude
must be ${\mathcal O}(p_{1}^{2})$. Since the Chern-Simons term is only ${\mathcal O}(p)$,
it would appear that the Abelian Chern-Simons term cannot be generated by radiative corrections.
However, this simple argument fails when there is no momentum input at the fermion triangles' third
vertices. If $p_{1}$ is always identically equal to $-p_{2}$, then the two transversality
conditions in (\ref{eq-MWard}) are redundant, and the matrix element only needs to be ${\cal O}(p_{1})$,
meaning a Chern-Simons term is actually allowed.

Without two independent Ward identities to be satisfied, the sum of the two triangle diagrams is
actually undetermined, because the diagrams are each naively linear divergent, and there is no unique
way to regulate them. However they are regulated, the divergent parts of the two diagrams will cancel,
producing a finite result. A specific regulator is often most conveniently expressed
in terms of a relationship between the loop momenta $k$ and $k'$ in the two diagrams.
If the amplitude really had needed to be transverse to two different photon momenta $p_{1}$
and $p_{2}$, it would have been necessary to choose $k'=k+3p_{1}$ and then (after Wick rotation) to perform
a spherically symmetric integration over $k$. This is why when the axial vector vertex represents a physical
$\pi^{0}$---which carries a nonvanishing momentum---the chiral anomaly gives a unique result
for the meson decay rate. However, when only a single transversality condition is imposed, it is possible
to have $k'=k+(3+\xi)p_{1}$ for any real value of $\xi$. While the induced Chern-Simons term vanishes for
$\xi=0$, with nonzero values of $\xi$ there is a $k_{AF}^{\mu}=-\xi q^{2}b^{\mu}/16\pi^{2}$ proportional to
$\xi$~\cite{ref-jackiw1}.
Each value of $\xi$ essentially defines a different quantum theory, all based on the same classical
Lagrangian. Lorentz- and CPT-violating QED is thus an example of a QFT with finite
but undetermined radiative corrections; some general characteristics of such theories are discussed
in~\cite{ref-jackiw3}.

With a momentum cutoff regulator, the shift in the integration by $\xi p_{1}$ produces a surface term,
which is allowed to be nonzero because the full diagram is divergent. This kind of surface term is
well known to create problems with gauge invariance. However, because of the presence of the $\gamma_{5}$
in the fermion loop, there must be a Levi-Civita tensor $\epsilon_{\mu\alpha\beta\gamma}$ in the
resulting radiative correction to the photon two-point function; and because of the total antisymmetry of
$\epsilon_{\mu\alpha\beta\gamma}$, the radiative correction (i.e.\ the induced
Chern-Simons term ${\cal L}_{AF}$) still obeys the Ward identity. This is
what ensures that the integrated action remains gauge invariant, even though gauge invariance is lost at
the level of the Lagrange density.

Since surface terms are involved, it seems like it might be possible to
avoid the Chern-Simons ambiguity by using a better regulator. However,
both Pauli-Villars and dimensional regularization---normally the best
choices when there are potential problems with maintaining gauge
invariance---reintroduce the ambiguity in other ways. The Pauli-Villars
method entails introducing additional families of fictitious heavy fermions,
whose contributions to the photon self-energy are subtractive. However, the
new fermions will posses their own $b^{\mu}$ terms, whose sizes are not
determined by the classical Lagrangian. In dimensional regularization,
there is no unique extension of $\gamma_{5}$ to $4-\epsilon$ dimensions,
and different extensions will produce different Chern-Simons terms. With
other regulation methods, the source of the radiative ambiguity may sometimes
be disguised, but the ambiguity always appears to be present somewhere.

Several specific nonzero values for the induced $k_{AF}^{\mu}$ were suggested in
the literature. These were typically based on various nonperturbative
arguments for how the momentum integrations in the two triangle diagrams
should be performed. However, any nonperturbative method should determine
the structure of the radiative corrections not just at ${\cal O}(b)$,but
also at ${\cal O}(b^{2})$. It turns out that any choice of regulator that
gives a specific nonzero coefficient for the Chern-Simons term at first
order also breaks gauge invariance by producing
a Lorentz-violating photon mass~\cite{ref-altschul8} term at second order
in $b^{\mu}$ (e.g~\cite{ref-altschul1,ref-altschul2,ref-filipe}). The one natural
exception is a regulator that produces a
vanishing $k_{AF}^{\mu}$; it is always possible to enforce the maximal degree of gauge
invariance at first order without spoiling gauge symmetry at higher
orders.

\subsection{Gravitational Theory}

Having pointed out all the key properties of the ambiguity in the Abelian
Chern-Simons term in Lorentz-violating QED, we now turn our attention to
the even trickier case of the gravitational Chern-Simons term.

The gravitational SME action includes the $b^{\mu}$ term in the form
\begin{equation}
\label{eq-Spsi}
S_{\psi}=\int d^{4}x\, ee^{\mu}\,_{a}\bar{\psi}\left(\frac{i}{2}\gamma^{a}
\overset{\text{\tiny$\leftrightarrow$}}{D}_{\mu}+
b_{\mu}\gamma^{a}\gamma_{5}\right)\psi.
\end{equation}
where the fermions are taken to be massless (purely for simplicity). The vierbein (tetrad) is $e^{\mu}\,_{a}$,
and its determinant is $e$. The coupling to gravitation occurs through $e^{\mu}\,_{a}$ and through
the gravitational covariant derivative, which is
\begin{equation}
D_{\mu}\psi=\partial_{\mu}\psi+\frac{1}{2}\omega_{\mu cd}\sigma^{cd}\psi,
\end{equation}
with the usual spin connection $\omega_{\mu}\,^{cd}$ including derivatives of
the vierbein.
Because of the required vierbein factors, (\ref{eq-Spsi}) is written as an integrated action $S$, although
for linearized gravity it would actually be sufficient to work with just a Lagrange density.
In a the linearized theory and in harmonic gauge, chosen for its simplicity and convenience, especially
when dealing with gravitational
anomalies~\cite{ref-alvarez}, the vierbein has a very simple representation in terms of the metric:
$e_{\mu a}=\eta_{\mu a}+\frac{1}{2}h_{\mu a}$ and $e^{\mu}\,_{a}=\eta^{\mu}\,_{a}-\frac{1}{2}h^{\mu}\,_{a}$.

\begin{figure}
\centering
\includegraphics{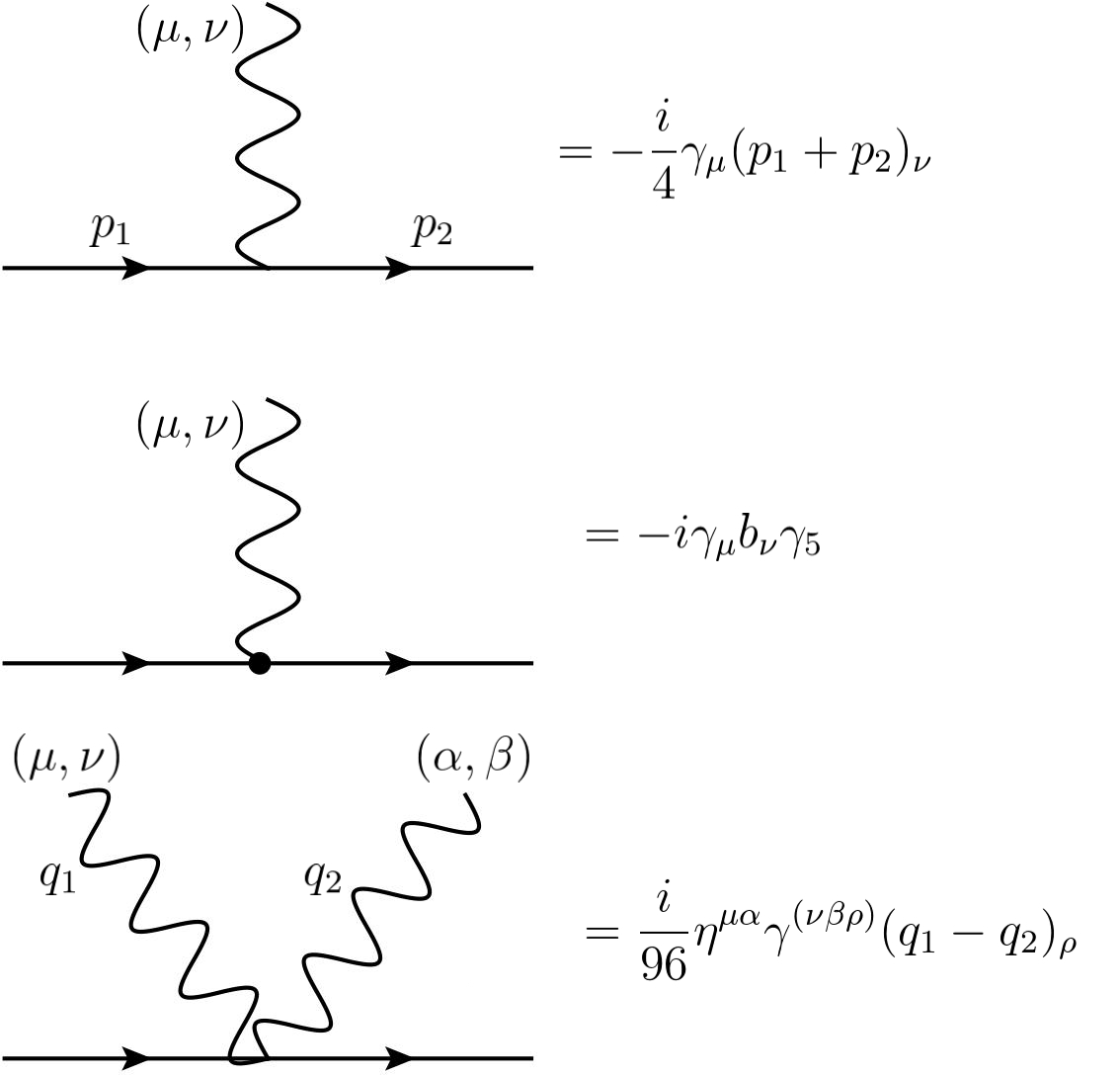}
\caption{Feynman rules for the fermion-graviton vertices in the presence
of $b^{\mu}$.
\label{fig-rules}}
\end{figure}

Neglecting $h^{\mu}\,_{\mu}$ interactions, which cannot contribute to the gravitational Chern-Simons
term, the linearized Lagrange density for the fermions coupled to gravity is
\begin{equation}
{\cal L}_{\psi}=\bar{\psi}\left\{\frac{i}{2}\left(\gamma^{\mu}-\frac{1}{2}h^{\mu\nu}\gamma_{\nu}\right)
\overset{\text{\tiny$\leftrightarrow$}}{\partial}_{\mu}
-h_{\mu\nu}\left[\frac{1}{2}b^{\mu}\gamma^{\nu}\gamma_{5}+
\frac{i}{96}(\partial_{\rho}h_{\alpha\beta})\eta^{\beta\nu}\gamma^{(\nu\beta\rho)}\right]
+\!\not\!b\gamma_{5}\right\}\psi.
\end{equation}
This expression involves the
antisymmetrized product of three distinct $\gamma$-matrices, $\gamma^{(\nu\beta\rho)}=
\gamma^{\nu}\gamma^{\beta}\gamma^{\rho}\pm({\rm all\, permutations})$.

The corresponding Feynman rules for the perturbative interactions of gravitons with
Lorentz-violating fermions are given in Figure~\ref{fig-rules}. There are the
usual vertices for single or paired gravitational excitations $h^{\mu\nu}$ interacting
with a fermion line, and there is also a new vertex in which $b^{\mu}$ appears.
The new vertex exists because of the presence of $b^{\mu}$ in the energy-momentum
tensor for the fermionic sector. However, it turns out that the new vertex does not
actually make any contribution to the radiatively-induced gravitational Chern-Simons term.
On the other hand, both the usual three-particle vertex and the four-particle vertex involving
$\gamma^{(\nu\beta\rho)}$ play potentially important roles.

\begin{figure}
\centering
\includegraphics[align=c]{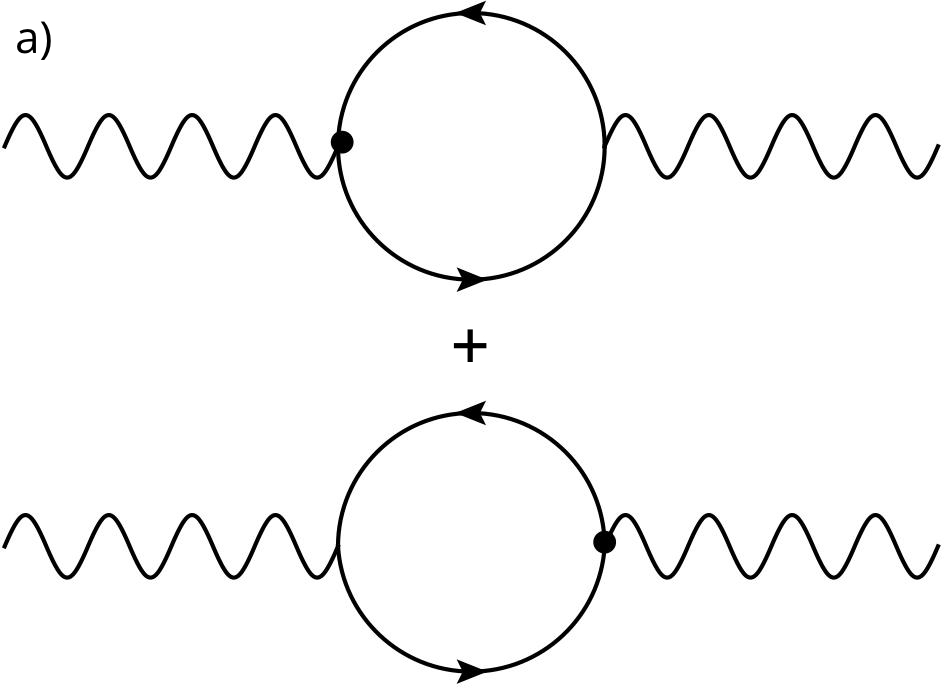}
\includegraphics[align=c]{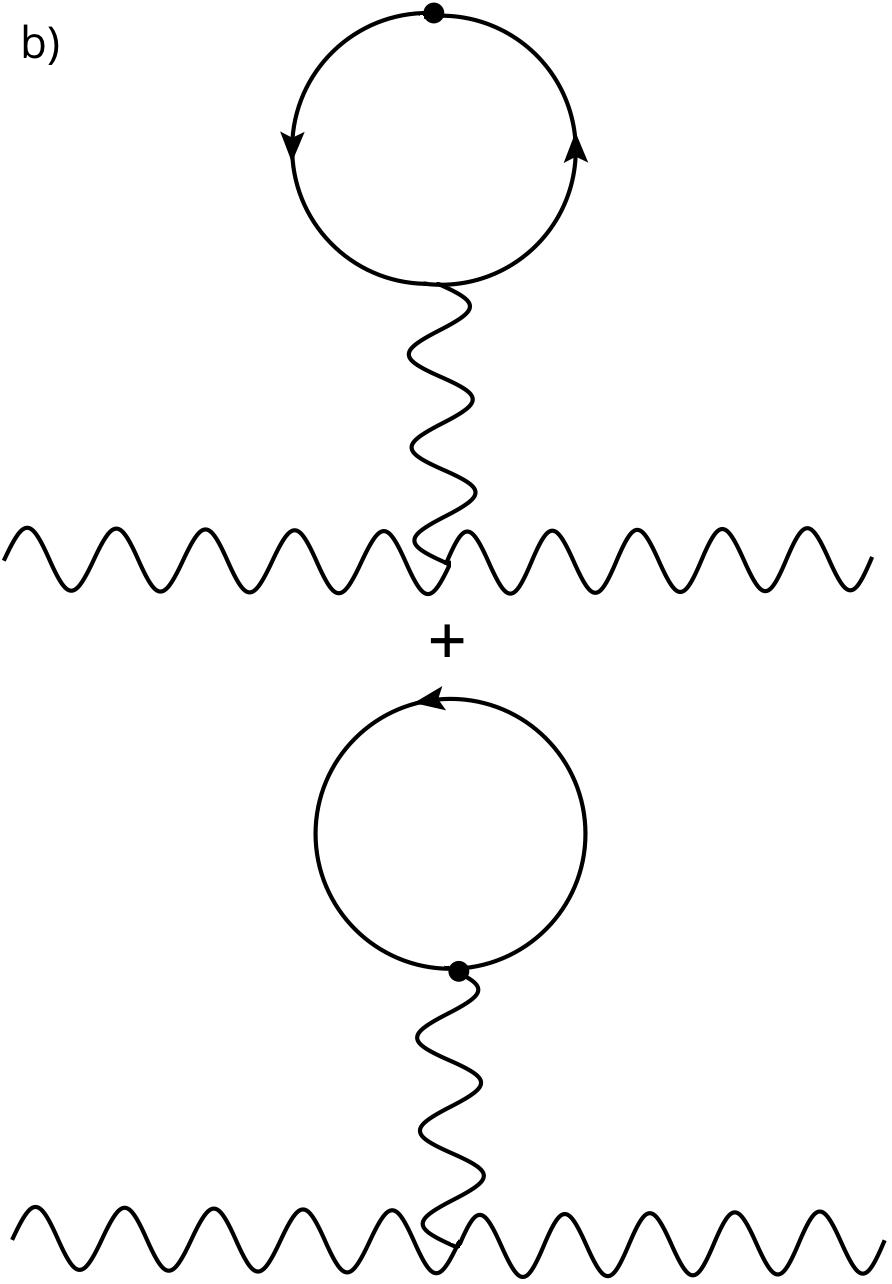}
\caption{Four diagrams that do not contribute to the radiatively induced gravitational
Chern-Simons term. (a) Diagrams with a $b^{\mu}$-modified fermion-photon vertex, which leads
to a term in which $b^{\mu}$ is contracted directly with one of the indices of the external
graviton $h^{\mu\nu}$. (b) Pure tadpole diagrams.
\label{fig-notcontrib}}
\end{figure}

All the two-point graviton diagrams derived from ${\cal L}_{\psi}$ that have a single fermion
loop are shown in Figures~\ref{fig-contribute}
and~\ref{fig-notcontrib}. However, only those in 
Figure~\ref{fig-contribute} can actually contribute to the Chern-Simons term (see below for details),
and we shall therefore concentrate our attention on those three diagrams.
Besides the presence of an additional diagram with a two-graviton vertex, there is another
way in which the gravitational radiative corrections are more complicated than those in the
Abelian theory.
Because the metric modes couple to the fermions' energy-momentum, there are additional
factors of the loop momentum appearing at the fermion-boson vertices. This gives the two triangle
diagrams in  Figure~\ref{fig-contribute}a a naive cubic degree of divergence. The new diagram with
the two-graviton vertex would also have a cubic divergence if the axial vector vertex could carry
a nonzero momentum. However, with only a strictly constant $b^{\mu}$ inserted into the fermion
propagator, the degree of divergence is reduced. In order to obtain
a finite final results for the induced $v^{\mu}$, both the cubic and linear divergences 
in the sum of the diagrams must be canceled.

The more elaborate cancelation is generally possible (and necessary in order to preserve
gravitational gauge invariance) because there are now three contributing diagrams---as
opposed to the just two that were present in the Abelian gauge theory. Note
that with a symmetric $k''$ integration, the diagram in Figure~\ref{fig-contribute}b actually gives
no net contribution to the Chern-Simons term; however, shifting the integration momentum by a
multiple of $p_{1}$ does yield a surface term with a linear divergence. The
calculation of the sum of the three diagrams
proceeds along much the same lines as in the original papers on the gravitational
contribution to the partially conserved axial current (PCAC)~\cite{ref-kimura,ref-delbourgo1,ref-delbourgo2}
(describing, for example, the potential rate of decay of a $\pi^{0}$ into two gravitons). The cubic
divergences in the triangle diagrams automatically cancel, because they have opposite momentum
routings and $\gamma_{5}$ is present. However, the cubic cancelation still leaves a residual linear
divergence. In the Abelian case, the residual finite term in the sum of the two triangle
diagrams was set by a $p_{1}$-dependent difference in the integration momenta $k$ and $k'$; here in
the gravitational calculation, the remaining linear divergence is set by the difference in $k$ and
$k'$. However, with a shift in the integration momentum $k''$ in the tadpole diagram, this linear
divergence may also be canceled.

These results, for both kinds of gauge theories,
can also be expressed in terms of the number of Ward identities that need to be satisfied.
When the axial current vertex carries an external momentum (as it does in $\pi^{0}$ decay), there
there are two independent Ward identities, because there are two different external momenta to
which the fermion loop matrix element must be transverse. Violations of gauge invariance must come
from divergent loop integrals. Imposing the first Ward identity always forces the strongest
naive divergence to vanish. However, there are still possible gauge symmetry violations coming
from surface terms associated with the Feynman diagram divergences. In the electromagnetic case (with
a naive linear divergence), there is one possible surface term, which may be adjusted to zero by
choosing $k-k'$. This enforces the second independent Ward identity.
In the gravitational version, there are two possible surface terms (a
linearly divergent one and a finite one), because the initial degree of divergence is cubic.
Again however, by suitable choices of both $k-k'$ and $k-k''$, both the surface terms may be
eliminated, again making the matrix element transverse to both boson momenta.

The situation is somewhat different when the axial vector vertex originates from
a constant Lorentz-violating background $b^{\mu}$. In that case, as we have noted, there is
only one Ward identity to enforce, because the momenta $p_{1}$ and $p_{2}$ of the external
gauge bosons are redundant. In the Abelian case, that means that $k-k'$ may be chosen to
be any multiple of $p_{1}$, yielding an undetermined Chern-Simons term. In the gravitational
theory, there is still one nontrivial affine condition relating $k$, $k'$, and $k''$ that
is needed to ensure the linear divergence cancels. However, this means once again that
there remains one undetermined parameter, and different choices of this parameter will produce
different values for the induced $v^{\mu}$.
Other regulators for the naively divergent diagrams introduce the ambiguity in other ways,
just as in the Abelian theory. It may seem natural, therefore, to conclude that the
coefficient of the induced gravitational Chern-Simons term should be entirely undetermined,
just as in the Abelian case.

Before continuing, we shall pause briefly to point out why the only possible contributions to the induced
Chern-Simons term actually come from the Feynman diagrams in Figure~\ref{fig-contribute}, even though
there are several other diagrams that can be constructed formally as contributions to the
graviton two-point function. The non-contributing diagrams are shown in Figure~\ref{fig-notcontrib}.
The two diagrams appearing in Figure~\ref{fig-notcontrib}a have the Lorentz violation
entering through the $b^{\mu}$-modified
interaction vertex from Figure~\ref{fig-rules}. However, any diagram with
a $b^{\mu}$ vertex could only give a contribution to the effective action with
the Lorentz index of $b^{\mu}$ directly contracted with one of the external $h^{\mu\nu}$
metric modes---and such a term is not of the Chern-Simons form.
Moreover, the other two diagrams in Figure~\ref{fig-notcontrib}b are not one-particle irreducible.
They have pure tadpole forms, with the intermediate graviton propagator necessarily carrying
vanishing momentum, so no information about the momentum of the external gravitons can reach
the divergent fermion loop. In any reasonable renormalization scheme, the sum of all the one-point
graviton tadpole diagrams will be set to vanish (so that the nonfluctuating part of the metric
$g^{\mu\nu}$ takes on its proper background value).

More details of the integrals, covering both the Abelian and QED cases, are given in the appendix.

\section{Resolving the Puzzle}

The results of the previous section
bring us back to the extremely puzzling point that while a $b^{\mu}$ term in the fermion
sector is Lorentz violating, any proportional $v^{\mu}$ that is induced in the linearized
gravity sector is not. It seems very counterintuitive that the quantum corrections
at first order in $b^{\mu}$ should somehow ``restore'' the broken Lorentz symmetry, at least in the
gravitational sector. There is no analogous symmetry restoration for the electromagnetic radiative
corrections; the $k_{AF}^{\mu}$ violates the same spacetime symmetries as a $b^{\mu}$ term. Doubly puzzling
is that while the gravitational Chern-Simons term avoids the Lorentz violation associated with $b^{\mu}$,
it still has the same discrete symmetries (and hence the same CPT violation) as the fermion sector
term.

However, this paradoxical---or at the least, extremely curious---behavior turns out to be an artifact
of having used an oversimplified description of Lorentz violation in conjunction with gravity.
In fact, the direct generation of a Lorentz-invariant radiative correction by a
Lorentz-violating term in the fermion sector does not actually occur.

We have repeatedly noted that when the axial vector vertex insertion in the fermion propagator can
carry a nonzero momentum, the number of constraints on a bosonic two-point function changes.
If momentum can be exchanged with the axial vector background, then the momenta of the
two attached bosons are not identical, and then there are two independent transverse Ward identities
that must hold. Then the argument that the radiative corrections to the boson propagator
must be at least ${\cal O}(p_{1}^{2})$ is entirely correct, and a Chern-Simons term is excluded. Notice
that insisting that the two separate Ward identities both hold when there is a net external momentum being
inserted into the loop diagrams is equivalent to demanding that the Fourier transform of the effective
Lagrange density be gauge invariant at the value of the external momentum in question. In fact, the only Fourier
component of the (electromagnetic or gravitational) Chern-Simons term that is invariant under gauge
transformations is the zero-momentum component---which is just the integrated action. If a stronger
form of gauge invariance---invariance of the density ${\cal L}$, rather than just $S$---is required,
then the coefficient $v^{\mu}$ in the gravitational sector must vanish.

It happens to be the case that
Lorentz violation in a metric theory of gravity must arise
spontaneously. Spontaneous Lorentz symmetry breaking is analogous to
other, more familiar, types of spontaneous symmetry breaking. A bosonic field
acquires a vacuum expectation value (vev), so that the vacuum state of the
theory does not respect all the symmetries of the underlying Lagrangian. If the
field with the vev has tensor indices, then the vev becomes a preferred
tensor in the vacuum. Couplings of other fields to the symmetry-breaking field
give rise to SME-type Lorentz-violating operators.

In a flat-space QFT like QED, Lorentz violation might arise
spontaneously, or it might be explicit. In the latter case, the fundamental
Lagrangian for the theory contains operators that violate Lorentz symmetry.
Either possibility is internally consistent, although what we know about
symmetry breaking in real physical systems may suggest that the spontaneous
symmetry breaking might be more elegant. In a gravitational theory, however,
matters are quite different. Only spontaneous symmetry breaking is possible;
gravity theories with explicit Lorentz breaking turn out to be mathematically
inconsistent. In a metric theory with explicit symmetry breaking, the Bianchi identities
cannot be satisfied, and the theory fails~\cite{ref-kost12}.

The qualitative reason for the inconsistency is actually rather simple. The basic premise of
a metric theory of gravitation is that test particles are moving along geodesics of
a background spacetime configuration. Two particles with equal mass, beginning
at the same point and moving with the same initial speeds, must follow exactly
the same trajectory. There is no room for the dynamics to depend on anything
else; there is no way to incorporate the spin and orientation dependences normally
associated with the motion of different species of particles in a theory with explicitly
broken rotation or boost symmetry. A preferred direction like $b^{\mu}$ cannot affect the motion of a
fermion if the fermion's motion is entirely determined by the spacetime geometry that
it is passing through. This argument holds even in a pure gravity theory,
because gravity actually provides its own test quanta. Pure gravity theories
have propagating gravitons, which in the linear theory are still effectively passing through a
background geometry.

This geometric obstruction has spurred some interest in studying Lorentz violation in more general Finsler
spacetimes. However, this work is still in its infancy; basic constructions,
such as of a scalar field action or a spinor bundle have not yet been demonstrated.
For the present purposes, we shall continue to
suppose that gravitation is described by a metric theory like general
relativity. The structure and consequences of a gravitational Chern-Simons-like term
in a Finsler geometry are matters far beyond the
current state of understanding.

The fact that Lorentz violation in a metric theory of gravity must be spontaneous
has definite phenomenalistic consequences. There must be additional fluctuating
modes in the theory, which affect the physical observables in both the purely 
gravitational sector~\cite{ref-bailey2} as well as with matter-gravity
couplings~\cite{ref-tasson1}.
The observation that the preferred $b^{\mu}$ is formed from the vacuum expectation value of 
vector-valued fields on the spacetime manifold leads to a number of interesting conclusions
and opens up new avenues of investigation. The vector field underlying $b^{\mu}$
may have a global structure related to the topology of the spacetime. Moreover, there will be
additional quantized excitations which are coupled to the theory's fermions
in the same way as $b^{\mu}$ itself.

However, what is important here is that the $b^{\mu}$
that appears in the Feynman rules in the gravitational theory cannot just be a
fixed background vector. Instead, it is accompanied by additional fluctuating degrees
of freedom. While the fluctuations themselves may be extremely small, the very fact
that they must be possible changes the conceptual nature of the $b^{\mu}$ vertices.
Because $b^{\mu}$ is the vev of a dynamical field, it has to be possible for there
to be momentum exchange between the $b^{\mu}$ vertex and the gravitational field.
When a nonzero momentum can enter the fermion loops at the $b^{\mu}$ insertion, there
are going to be two nontrivial Ward identities, and the effective Lagrange density
itself---not just the integrated effective action---must be gauge invariant. This
returns the theory to the original situation that was studied in the context of PCAC,
in which the form of the radiative corrections is completely fixed. For the
coefficient of the induced gravitational Chern-Simons term, the resulting unambiguous
value is zero.

Thus the highly peculiar behavior of the radiative corrections---that there could be a
Lorentz-invariant correction that is linear in the Lorentz-violating parameter
$b^{\mu}$---has thus been avoided. The reason for this is that
the arbitrariness of the Chern-Simons-type radiative corrections
only exists when the Lorentz violation in the theory is explicit---which itself
may be a rather unexpected result, although not a potentially paradoxical one.
In fact, we may take the argument one step further and note that the when
Lorentz-violating QED is studied in the (realistic) context of a background spacetime
governed by general relativity, the $b^{\mu}$ in the fermion sector still has to be
just one piece of a dynamical field. This means that the radiative-induced Chern-Simons
term for the Abelian theory is also zero!

We have reached this level of understanding by drawing together earlier conclusions
about the mathematical structures of different kinds of Lorentz-violating field
theories. This further reinforces the observation that when basic symmetries such as
Lorentz symmetry or CPT are broken there may be some fairly subtle effects,
qualitatively unlike those seen in more symmetric models---especially in regard
to quantum corrections.

\appendix

\section*{Appendix: Fermion Loop Integrals}

To see how the cancelation between divergent terms works in the Lorentz-violating
theories (both electromagnetic and gravitational) it is useful to inspect the
loop integrals involved explicitly. For the QED case, the integration structure
has been explored by a variety of methods in~\cite{ref-jackiw1,ref-victoria1,ref-chung2},
including the nonanalytic dependences on both $b^{\mu}$ and $p^{\mu}$. However,
the key features for our analysis appear at leading order in both these quantities,
so we shall limit ourselves to that order. At first order in $b^{\mu}$, the
relevant terms in the photon self-energy are given by
\begin{eqnarray}
\Pi^{\mu\nu}(p) & = & -iq^{2}\int\frac{d^{4}k}{(2\pi)^{4}}{\rm tr}\left\{
\gamma^{\mu}\frac{i}{\!\not k-m}\gamma^{\nu}\frac{i}{\!\not k\,+\!\not\! p-m}\left(i
\!\not\!b\gamma_{5}\right)\frac{i}{\!\not k\,+\!\not\! p-m}\right. \nonumber\\
& & +\left.\gamma^{\nu}\frac{i}{\!\not k+\zeta\!\not\! p-m}\gamma^{\mu}\frac{i}{\!\not k+(\zeta+1)\!\not\! p-m}
\left(i\!\not\!b\gamma_{5}\right)\frac{i}{\!\not k+(\zeta+1)\!\not\! p-m}\right\}.
\label{eq-Pi-QED}
\end{eqnarray}
The free parameter $\zeta$ describes the arbitrary momentum routing difference between the two
contributing diagrams.

The $p$-independent contribution to this self-energy is simply
\begin{equation}
\label{eq-Pi0}
\Pi^{\mu\nu}(0)=-iq^{2}b^{\alpha}\int\frac{d^{4}k}{(2\pi)^{4}}{\rm tr}\left\{
\gamma^{\mu}\frac{\!\not k+m}{k^{2}-m^{2}}\gamma^{\nu}\frac{\!\not k+m}{k^{2}-m^{2}}
\gamma_{\alpha}\gamma_{5}\frac{\!\not k+m}{k^{2}-m^{2}}+(\mu\leftrightarrow\nu)\right\}.
\end{equation}
This can be shown to vanish using explicit Dirac algebra calculations. However,
following~\cite{ref-jackiw1}, it is simpler to notice that the contraction of the integral in (\ref{eq-Pi0})
with $b^{\alpha}$ is a two-index object that is
proportional to the Levi-Civita tensor $\epsilon^{\alpha\beta\gamma\delta}b_{\alpha}$,
yet independent of $p$; no such object can be constructed, so the
integral must vanish. Note that the involvement of the $\gamma_{5}$ (which is defined in
terms of the Levi-Civita tensor) plays a key role in the cancelation; virtual fermions with
opposite helicities contribute oppositely to the naively linearly divergent term.

As already noted, the term linear in the momentum vanishes for $\zeta=3+\xi=3$, which is 
the choice of momentum routing in the usual calculation of the QED chiral anomaly. The
nonzero contribution then comes entirely from a surface term, which is given by the
shifting the loop momentum in one diagram by $\xi p$,
\begin{eqnarray}
\Pi^{\mu\nu}(p) & = & q^{2}b^{\alpha}\int\frac{d^{4}k}{(2\pi)^{4}}{\rm tr}\left\{
\gamma^{\mu}\frac{1}{\!\not k-m}\gamma_{\alpha}\gamma_{5}\frac{1}{\!\not k-m}
\gamma^{\nu}\frac{1}{\!\not k\,+\xi\!\not\! p-m}\right. \nonumber\\
& & +\left.\gamma^{\mu}\frac{1}{\!\not k\,+\xi\!\not\! p-m}\gamma_{\alpha}\gamma_{5}\frac{1}{\!\not k\,+\xi\!\not\! p-m}
\gamma^{\nu}\frac{1}{\!\not k-m}\right\} \\
& = & q^{2}b^{\alpha}\int\frac{d^{4}k}{(2\pi)^{4}}{\rm tr}\left\{
\gamma^{\mu}\frac{\!\not k+m}{k^{2}-m^{2}}\gamma_{\alpha}\gamma_{5}\frac{\!\not k+m}{k^{2}-m^{2}}
\gamma^{\nu}\frac{\!\not k+m}{k^{2}-m^{2}}\xi\!\not\! p\frac{\!\not k+m}{k^{2}-m^{2}}\right. \nonumber\\
& & +\left.\gamma^{\mu}\frac{\!\not k+m}{k^{2}-m^{2}}\xi\!\not\! p\frac{\!\not k+m}{k^{2}-m^{2}}
\gamma_{\alpha}\gamma_{5}\frac{\!\not k+m}{k^{2}-m^{2}}\xi\!\not\! p\frac{\!\not k+m}{k^{2}-m^{2}}
\gamma^{\nu}\frac{\!\not k+m}{k^{2}-m^{2}}\right\}.
\end{eqnarray}
The remaining divergence [coming from the terms in integrand that are proportional to
$k^{4}/(k^{2}+m^{2})^{4}$] in this surface term expression cancels~\cite{ref-chung2},
leaving a finite expression; the straightforward result is, at leading order in $p$,
\begin{equation}
\Pi^{\mu\nu}(p)=-\frac{\xi q^{2}}{8\pi^{2}}\epsilon^{\mu\nu\alpha\beta}b_{\alpha}p_{\beta},
\end{equation}
corresponding to the ambiguous $k_{AF}^{\mu}$.

The cancelations function similarly for the gravitational Chern-Simons term, but they are more
elaborate, since the naive degree of divergences of the integrals involved are greater. However,
it is not excessively difficult to get the potentially nonzero contribution to the
the radiatively generated $v^{\mu}$ in a form that closely mirror the results in the QED theory.
The version of the calculation that arises in the PCAC context (with the axial vector insertion
carrying nonzero momentum) is given in detail in~\cite{ref-kimura}, so we will concentrate on
the case where the $b^{\mu}$ carries no momentum---the situation out of which a potentially
ambiguous radiative correction could arise.

We will begin with the fermion tadpole diagram shown in Figure~\ref{fig-contribute}b, which does
not contribute directly to the Chern-Simons term when the $k$-integration momentum is done
symmetrically. If the external gravitons carry momentum $p$, and have polarization indices
$(\mu,\nu)$ and $(\alpha,\beta)$, respectively, the Feynman rules give a self-energy tensor
\begin{equation}
\label{eq-Pi1b}
\Pi^{\mu\nu\alpha\beta}_{1b}(p)=-\frac{i}{96}\int\frac{d^{4}k}{(2\pi)^{4}}{\rm tr}\left\{
\eta^{\mu\alpha}\gamma^{(\nu\beta\rho)}(2p_{\rho})\frac{i}{\!\not k}
\left(i\!\not\!b\gamma_{5}\right)\frac{i}{\!\not k}\right\},
\end{equation}
recalling that we are using massless fermions for simplicity. With the symmetric integration,
there is no contribution in~(\ref{eq-Pi1b})
to the gravitational Chern-Simons term, which is ${\cal O}(p^{3})$. What is more, the massless
integral, while superficially quadratically divergent, actually vanishes if the integration
is done by dimensional regularization and evaluated at $d=4$. Note that the trace in~(\ref{eq-Pi1b}), combined
with symmetric integration over $k$, gives an overall expression proportional to
$\eta^{\mu\alpha}\epsilon^{\nu\beta\rho\delta}p_{\rho}b_{\delta}$, so this term, on its
own, would (if it were not vanishing) satisfy the Ward identities for transversality in
the indices $\nu$ and $\beta$.

The important part of the
potentially ambiguous radiative correction in the gravitational theory really comes (as in the
Abelian gauge theory) from the two triangle diagrams from Figure~\ref{fig-contribute}a. The sum of
the diagrams looks similar to what appears in the QED version, but there is an additional momentum
factor at each fermion-boson vertex. The sum of the two triangles, again with an undetermined
difference $\zeta p$ between the integration momenta, is
\begin{eqnarray}
\Pi^{\mu\nu\alpha\beta}_{1a}(p) & = & \frac{i}{16}\int\frac{d^{4}k}{(2\pi)^{4}}{\rm tr}\left\{
\gamma^{\mu}(2k+p)^{\nu}\frac{i}{\!\not k}\gamma^{\alpha}(2k+p)^{\beta}\frac{i}{\!\not k\,+\!\not\! p}\left(i
\!\not\!b\gamma_{5}\right)\frac{i}{\!\not k\,+\!\not\! p}\right. \nonumber\\
& & +\gamma^{\alpha}[2k+(2\zeta+1)p]^{\beta}\frac{i}{\!\not k+\zeta\!\not\! p}\gamma^{\mu}
[2k+(2\zeta+1)p]^{\nu} \nonumber\\
& & \times\left.\frac{i}{\!\not k+(\zeta+1)\!\not\! p}
\left(i\!\not\!b\gamma_{5}\right)\frac{i}{\!\not k+(\zeta+1)\!\not\! p}\right\}.
\end{eqnarray}
As in~(\ref{eq-Pi0}), the most divergent contribution $\Pi^{\mu\nu\alpha\beta}_{1a}(0)$
(with a cubic divergence in this case) must vanish, since it has an impossible tensor
structure---symmetric in $(\mu,\nu)$ and $(\alpha,\beta)$ and also
proportional to the Levi-Civita tensor, but independent of $p$;
this necessarily means $\Pi^{\mu\nu\alpha\beta}_{1a}(0)=0$.

The quadratically divergent part of $\Pi^{\mu\nu\alpha\beta}_{1a}(p)$ then
automatically vanishes in the
same way as the quadratic divergence in $\Pi^{\mu\nu\alpha\beta}_{1b}(p)$, using
Wick-rotated dimensional regularization at $d=4$.

What remains is a difference of two linearly divergent integrals, as in the QED case. Explicit
calculation of the full integrals is extraordinarily tedious. However, it is possible to
verify that $\zeta$ plays the same role in parameterizing the ambiguity in the gravitational theory
as in the Abelian theory in a relatively straightforward way. This is done by
differentiating $\Pi^{\mu\nu\alpha\beta}_{1a}(p)$ with respect to $\zeta$, giving
\begin{eqnarray}
\label{eq-Pi-deriv}
\left.\frac{d\Pi^{\mu\nu\alpha\beta}_{1a}}{d\zeta}\right|_{\zeta=0} & = &
\frac{i}{16}\int\frac{d^{4}k}{(2\pi)^{4}}{\rm tr}\left\{
\gamma^{\alpha}(2p)^{\beta}\frac{1}{\!\not k}\gamma^{\mu}
(2k+p)^{\nu}\frac{1}{\!\not k\,+\!\not\! p}
\!\not\!b\gamma_{5}\frac{1}{\!\not k\,+\!\not\! p}\right. \\
& & +\,\gamma^{\alpha}(2k+p)^{\beta}\frac{1}{\!\not k}(-\!\not\! p)\frac{1}{\!\not k}
\gamma^{\mu}(2k+p)^{\nu}\frac{1}{\!\not k\,+\!\not\! p}
\!\not\!b\gamma_{5}\frac{1}{\!\not k\,+\!\not\! p} \nonumber\\
& & +\,\gamma^{\alpha}(2k+p)^{\beta}\frac{1}{\!\not k}\gamma^{\mu}
(2p)^{\nu}\frac{1}{\!\not k\,+\!\not\! p}
\!\not\!b\gamma_{5}\frac{1}{\!\not k\,+\!\not\! p}
+\gamma^{\alpha}(2k+p)^{\beta}\frac{1}{\!\not k}\gamma^{\mu}
(2k+p)^{\nu}
\nonumber\\
& & \times\left.\frac{1}{\!\not k\,+\!\not\! p}\left[
(-\!\not\! p)\frac{1}{\!\not k\,+\!\not\! p}\!\not\!b\gamma_{5}
+\!\not\!b\gamma_{5}\frac{1}{\!\not k\,+\!\not\! p}(-\!\not\! p)
\right]\frac{1}{\!\not k\,+\!\not\! p}\right\}. \nonumber
\end{eqnarray}
Taking into account the fact that the parts of this expression which come from
terms with quadratic and higher divergences must vanish, (\ref{eq-Pi-deriv}) simplifies to
\begin{eqnarray}
\left.\frac{d\Pi^{\mu\nu\alpha\beta}_{1a}}{d\zeta}\right|_{\zeta=0} & = &
-\frac{i}{16}\int\frac{d^{4}k}{(2\pi)^{4}}{\rm tr}\left\{
\gamma^{\alpha}p^{\beta}\frac{1}{\!\not k}\!\not\! p\frac{1}{\!\not k}
\gamma^{\mu}p^{\nu}\frac{1}{\!\not k\,+\!\not\! p}
\!\not\!b\gamma_{5}\frac{1}{\!\not k\,+\!\not\! p}\right. \nonumber\\
& & +\,\left.\gamma^{\alpha}p^{\beta}\frac{1}{\!\not k}\gamma^{\mu}
p^{\nu}\frac{1}{\!\not k\,+\!\not\! p}\left[
\!\not\! p\frac{1}{\!\not k\,+\!\not\! p}\!\not\!b\gamma_{5}
+\!\not\!b\gamma_{5}\frac{1}{\!\not k\,+\!\not\! p}\!\not\! p
\right]\frac{1}{\!\not k\,+\!\not\! p}\right\}.
\label{eq-Pi-deriv-2}
\end{eqnarray}
This has the same functional form as the equivalent derivative of~(\ref{eq-Pi-QED})
in the massless limit,
$d\Pi^{\mu\nu}/d\zeta|_{\zeta=0}$. There are two extra powers of the momentum
in~(\ref{eq-Pi-deriv-2}), which just correspond to the additional derivatives in the
gravitational Chern-Simons term. Since a linear shift in $\zeta$ in the Abelian theory
creates a proportional shift in the induced $k_{AF}^{\mu}$, a similar change in the
momentum routing in the gravitational theory would produce an equivalent linear shift
in $v^{\mu}$.

\end{document}